# A Majority Rule Philosophy for Instant Runoff Voting

Ross Hyman[a]*, Deb Otis[b], Seamus Allen[b], and Greg Dennis[c]

[a]*Research Computing Center, University of Chicago, Chicago, IL, USA*
[b]*FairVote, Silver Spring, MD, USA*
[c]*Voter Choice Massachusetts, Boston, MA, USA*

*rhyman@uchicago.edu

# A Majority Rule Philosophy for Instant Runoff Voting

We present the *core support criterion,* a voting criterion satisfied by Instant Runoff Voting (IRV) that is analogous to the Condorcet criterion but reflective of a different majority rule philosophy. Condorcet methods can be thought of as conducting elections between each pair of candidates, counting all ballots to determine the winner of each pair-election. IRV can also be thought of as conducting elections between all pairs of candidates but for each pair-election only counting ballots from voters who do not prefer another major candidate (as determined self-consistently from the IRV social ranking) to the two candidates in contention. The appropriateness of including all ballots or a subset of ballots for a pair-election, depends on whether the society deems the entire or a selected ballot set in compliance with freedom of association (which implies freedom of non-association) for a given pair election. Arguments based on freedom of association rely on more information about an electorate than can be learned from ranked ballots alone. We present a freedom-of-association based argument to explain why IRV may be preferable to Condorcet in some circumstances, including the 2022 Alaska special congressional election, based on the political context of that election.

**Statements and Declarations**
The authors report there are no competing interests to declare.

## I. Introduction

There are many ways to aggregate voter preferences in order to determine the winner of an election. The choice of a voting method depends upon the priorities of the society because no voting method can satisfy the full list of supposedly-desirable criteria. This paper compares Instant Runoff Voting (IRV) and Condorcet voting methods. IRV and Condorcet methods typically select the same winner (Graham-Squire et al, 2023; Stephanopoulos, 2024) but in the cases where they differ, the winning candidates reflect different philosophies of majority rule. Hill (2022) distinguishes between core support and broad support in their comparison of IRV and Condorcet. In this paper we quantify and generalize this distinction.

To compare IRV to Condorcet we present the *core support criterion* satisfied by IRV that is directly comparable to the Condorcet criterion. The two criteria are expressions of conflicting majority rule philosophies. They both concern elections between all pairs of candidates in an election of many candidates. While Condorcet methods count all ballots when determining the winner of each pair election, IRV counts ballots from voters who do not prefer another major candidate (as determined self-consistently from the IRV social ranking). While the exclusion of some ballots in determining the social order of any two candidates is a violation of the Condorcet majority rule philosophy, most societies incorporate the concept of freedom of association (which implies freedom of non-association) into their majority rule philosophy. Examples include political primary elections and caucus elections in parliaments, where only people inside the party or caucus have a vote. Therefore the appropriateness of including all ballots or a subset of ballots for a pair-election, depends on whether the society deems the entire or a selected ballot set in compliance with freedom of association for a given pair election. Arguments based on freedom of association rely on more information about an electorate than can be learned from ranked ballots alone. We present a freedom-of-association based argument to explain why IRV may be preferable to Condorcet in some circumstances, including the 2022 Alaska special congressional election, based on the political context of that election.

In Section II we introduce the 2022 Alaska special congressional ranked-ballot election which we use as an example throughout the paper. In Section III we introduce Instant Runoff Voting and the core support criterion that it satisfies. In section IV we describe how the political context of the Alaska election is consistent with the core support criterion. In Section V we describe how the core support criterion is incompatible with independence of irrelevant alternatives (IIA) and monotonicity. In Section VI we show how the Condorcet method satisfies IIA and monotonicity because it satisfies a broad support criterion rather than core support. For the 2022 Alaska special congressional election, The Condorcet winner has the most broad support and the least core support, while the IRV winner has the most core support and the second highest (and next to lowest) broad support.

## II. The 2022 Alaska Special Congressional Election
The voting methods described in this paper rely on **ranked ballots**, in which voters rank candidates in order of preference. Voters need not rank all candidates.

As an example of ranked ballots, we consider the 2022 Alaska special congressional election. The ballot totals compiled by Graham-Squire et al (2022) are shown in Table 1.

| Number of Voters | 27053 | 15467 | 11290 | 34049 | 3652 | 21272 | 47407 | 4645 | 23747 |
|---|---|---|---|---|---|---|---|---|---|
| Ist choice | Begich | Begich | Begich | Palin | Palin | Palin | Peltola | Peltola | Peltola |
| 2nd choice | Palin | Peltola | | Begich | Peltola | | Begich | Palin | |
| 3rd choice | Peltola | Palin | | Peltola | Begich | | Palin | Begich | |

Table 1: Ballot totals compiled by Graham-Squire et al (2022) for the 2022 Alaska special congressional election.

This ballot set is exceptionally rare (Graham-Squire et al 2023, Stephanopoulos 2024) for a real election because the IRV and Condorcet winners are different. The total number of recorded votes cast was 188,582. The candidates were Mary Peltola, a Democrat with 75,799 first place votes (40.19% of the total), Nick Begich, a Republican with 53,810 first place votes (28.53% of the total) and Sarah Palin, also a Republican but ideologically distinct from and to the right of Begich with 58,973 first place votes (31.27% of the vote). No candidate had a majority of 1st choice votes.

The voting methods described in this paper produce a complete **social ranking,** a transitive ranking of all of the candidates, from this ballot set. The topmost candidate is the winner of the election. For the plurality voting method, where only first choices are recorded, the social ranking is Peltola > Palin > Begich.

### III. The Instant Runoff Voting method and the Core Support Criterion
**Instant Runoff Voting (IRV)** is a ranked voting method in which the candidate ranking highest on the fewest number of ballots is excluded from ballots, placed in the lowest unoccupied position of the social ranking and this procedure is repeated until all candidates have been socially ranked. The highest ranked candidate is the IRV winner of the election (Tideman, 2006).

Example: Begich has the smallest number of first place votes and is placed third in the social ranking. Excluding Begich, Palin has fewer votes (58,973 ballots that rank Palin first plus 27,053 that rank Begich first and Palin second equals 86,026) than Peltola (75,799 ballots that rank Peltola first and 15,467 that rank Begich first and Peltola second equals 91,266) and is placed second in the social ranking. Peltola is placed first in the social ranking. The IRV social ranking is Peltola > Palin > Begich. Peltola wins the election. The IRV social ranking is the same as the plurality social ranking, but that is not always the case.

IRV is based on the idea that the candidate with the fewest votes should not win and this idea is applied repeatedly as each last place candidate is excluded. We now proceed to make the description of IRV comparable to voting methods that compare pairs of candidates, at the expense of making that description much more complicated.

Given any candidate, *Y*, we define a **minor candidate** relative to *Y* as any candidate lower in the social ranking  than *Y*. We define a **major candidate** relative to *Y* as all other candidates. In particular, candidate Y is major relative to Y. Major and minor candidates are always defined relative to a candidate. With the IRV social ranking, Peltola is a major candidate relative to Peltola, Palin, and Begich. Palin is a minor candidate relative to Peltola and a major candidate relative to Palin, and Begich. Begich is a minor candidate relative to Peltola and Palin and a major candidate relative to Begich.

When considering two candidates *X* and *Y* and an appropriate set of ballots, which could be all of the ballots or a subset of the ballots consistent with society's understanding of freedom of association, adherence to **majority rule** implies that if X is higher in the social ranking than Y then there are more ballots in the relevant subset ranking X higher than Y than ballots ranking Y higher than X. In this paper we are only considering voting methods that adhere to majority rule.

The **core support** for candidate X relative to candidate Y, where X is a major candidate relative to Y, includes all of the voters who rank X highest of all the major candidates relative to Y.

Examples:  The core support for Peltola relative to Begich includes only those ballots that rank Peltola first. Ballots that rank Palin first are not included even if they rank Peltola above Begich, because Palin is a major candidate relative to Begich. Likewise, the core support for Begich relative to Begich includes only those ballots that rank Begich first and do not include any ballots that rank Peltola or Palin first. The core support for Peltola relative to Palin includes all ballots that rank Peltola above Palin, including those that rank Begich first.

The **core support criterion** is satisfied, self-consistently, if the relative social order between any two candidates X and Y, with X higher in the social ranking than Y is determined from majority rule using only those ballots included in the core support for X relative to Y and Y relative to Y. That is, the ballots included do not rate a major candidate relative to Y higher in its ranking than the higher of X and Y. The ballots can rank minor candidates higher than the higher of X and Y. The motivation for the core support criterion is that voters should be able to rank minor candidate favorites without harming their influence on the choice of lower ranked major candidates but, in keeping with the principle of freedom of association, voters who prefer candidate Z who is a major candidate relative to Y, to X and Y, and is therefore not in the core support for X or Y relative to Y, do not get to have influence on the pair elections between X and Y.

The IRV social ranking Peltola > Palin > Begich satisfies the core support criterion. All ballots are accessed to determine the relative social ranking of Peltola and Palin because, besides Peltola and Palin there are no other major candidates relative to Palin.  Using all of the ballots, Peltola is preferred to Palin 91,266 to 86,026 with 11,290 ballots not stating a preference. All ballots that do not rank Palin first are used to determine the relative social ranking of Peltola and Begich because Palin is major to Begich. Using the allowed ballots, Peltola is preferred to Begich 75,799 to 53,810. All ballots that do not rank Peltola first are used to determine the relative social ranking of Palin and Begich because Peltola is major to Begich. Using the allowed ballots, Palin is preferred to Begich 58,973 to 53,810.

IRV is the *only* voting method that adheres to the core support criterion. To show this, consider the lowest candidate, Y, in the social ranking. The social ranking between Y and every other candidate X is determined by majority rule from those ballots that rank X or Y first as there are no minor candidates relative to Y. Therefore Y must have fewer first place votes than any other candidate, which is the same requirement for the lowest socially ranked candidate from the IRV method. As minor candidates defined in the core support criterion are treated the same as excluded candidates for the IRV method, major candidates defined in the core support criterion are treated the same as remaining candidates for the IRV method, and having fewer votes than any other major candidate is the same as having the fewest votes amongst the remaining candidates. IRV and the core support criterion are equivalent.

## IV. Political Context

IRV does not allow the lower ranked preferences of Peltola voters to influence the social order between Begich and Palin. This is consistent with a majority rule philosophy that finds it appropriate for Republican Party members to choose a candidate without interference from Democratic Party voters.

Likewise, IRV does not allow the lower-ranked preferences of Palin voters to influence the social order between Begich and Palin. This is consistent with a majority rule philosophy that finds it appropriate for voters who oppose Trump to choose a candidate without interference from voters who support Trump.

Lastly, IRV allows the lower ranked preferences of Begich voters to influence the social order between Peltola and Palin because Begich is a minor candidate relative to Peltola and Palin and perceived as more willing to actually support one of the other candidates. This is consistent with the fact that a higher percentage of Begich voters (79.02%) ranked a second candidate than voters who ranked Palin (63.93%) or Peltola (68.67%) first.

## V. Independence of Irrelevant Alternatives and Monotonicity

Kenneth Arrow's **impossibility theorem** (Arrow, 1963), later expanded by the Gibbard-Satterthwaite theorem (Gibbard, 1973; Satterthwaite, 1975), shows that no ranked voting method can satisfy all of a certain set of criteria identified as desirable. In particular IRV does not comply with Independence of Irrelevant Alternatives and monotonicity (Fishbum & Brams, 1983).

The **independence of irrelevant alternatives** (IIA) criterion states that the social preferences between two candidates should depend only on the individual relative preferences between those candidates (Arrow, 1963).

IRV is incompatible with IIA. Palin is far from irrelevant to the relative social ranking of Peltola and Begich because it is a major candidate relative to Begich so Palin supporters do not get to have influence over the choice of Peltola and Begich. However, if Palin is removed, those voters do get to choose between Peltola and Begich and Begich wins. The relevance of the IIA criterion depends on the majority rule philosophy concerning core support and freedom of association.

The **monotonicity criterion** tests whether it is possible to prevent the election of a candidate by ranking them higher on some of the ballots, or elect a candidate by ranking them lower on some of the ballots (Woodall, 1997).

IRV violates monotonicity. If 5,164 Palin voters change their first place vote to Peltola, that would decrease Palin's first place votes to below Begich. Palin will be eliminated and Begich wins the election between Begich and Peltola using majority rule on all of the ballots. This is a violation of monotonicity because the raising of Peltola on some ballots caused Peltola to lose the election.

The possibility of monotonicity violation is necessary, if it is deemed desirable for supporters of a minor candidate to have influence over the decision between two major candidates but not desirable for supporters of another major candidate to have influence. This violates monotonicity because increasing the vote total for the minor candidate so that it becomes a major candidate decreases the influence of those ballots. The necessity of upholding or violating the monotonicity criterion therefore depends on the value society places on core support and freedom of association.

## VI. Broad Support and the Condorcet Method
Many voting methods theorists consider monotonicity and IIA failure to be serious flaws in a voting method (Foley & Maskin, 2022; Ornstein & Norman, 2014) but this is not a universal opinion (Austen-Smith & Banks, 1991).

The criteria can be complied with by changing the voting method to determine the relative ranking of pairs of candidates using all of the ballots instead of just the core support ballots.

The **broad support** for candidate X relative to Y includes all of the voters who rank X above Y on all ballots. Ballots that are included in the broad support for X relative to Y but that are not included in the core support for X relative to Y, are cast from voters who prefer another major candidate relative to Y. For example, the broad support for Peltola relative to Begich includes all ballots that rank Peltola above Begich, including the core support ballots that ranked Peltola first and the non-core support ballots that ranked Palin first and Peltola second.

The **broad support criterion** is satisfied, if the relative social order between any two candidates X and Y is determined from  majority rule using all ballots that express a preference for X or Y. Broad support combines IIA and majority rule.

Voting methods that adhere to the broad support criterion are called **Condorcet voting methods**. Condorcet voting methods have long enjoyed popularity among voting methods enthusiasts (Tideman, 2006). When a transitive Condorcet social ranking exists, Condorcet methods satisfy many evaluative criteria, including independence of irrelevant alternatives and monotonicity. Green-Armytage (2011), Foley & Maskin (2022), and Ornstein & Norman (2014) argue for an interpretation of majority rule that requires the election of the Condorcet winner if one exists.

When comparing IRV and Condorcet, we see that both IRV and Condorcet agree that when determining the relative social order between two candidates, voters who do not prefer other major candidates relative to the pair should have their ballots counted for that pair-election. IRV and Condorcet differ on whether to allow voters who do prefer other major candidates relative to the pair to have their ballots counted for that pair election.

Using all of the ballots for each pair election, Begich is preferred to Peltola 87,859 to 79,451 with 21,272 not stating a preference. Begich is preferred to Palin 101,217 to 63618 with 23747 not stating a preference. Peltola is preferred to Palin 91,266 to 86,026 with 11290 not stating a preference. The Condorcet social order is Begich > Peltola >Palin, with Begich winning the election.

Begich is the first place finisher in the Condorcet social ranking. He has the most broad support, beating every other candidate in majority rule pair-elections counting all ballots. Begich is also the third place finisher in the IRV social ranking. He loses to every candidate in majority rule pair-elections counting only core support ballots relative to Begich. Peltola is the first place finisher in the IRV social ranking. She has the most core support but is in second place in the broad support Condorcet ranking. This is a general result whenever there are three candidates and the Condorcet and IRV winners differ.

## VII. Conclusions

IRV satisfies a majority rule philosophy, in which the relative social order between any two candidates is determined by counting only ballots from those voters who do not prefer another major candidate, while ignoring all minor candidates, in which major and minor are determined by the social order of the voting method and relative to the lower of the two candidates under consideration.

IRV is incompatible with the  independence of irrelevant alternatives because IIA demands that the relative social ranking between two candidates should depend on the relative social ranking of these candidates on *all* ballots. In addition, it is incompatible with monotonicity because monotonicity demands that if supporters of minor candidates have influence over lower ranked candidates then supporters of major candidates should have equal or greater influence, while IRV does not grant influence over the social ranking of two candidates by voters who prefer a different major candidate.  IRV can be preferable to Condorcet for elections in which allowing supporters of a major candidate to have influence over the relative social ranking between other major candidates is deemed inappropriate.